# The Holmberg effect of Main galaxy pairs of the SDSS Data Release 4 (SDSS4)

**Deng Xin-Fa** [1]  **Ma Xin-Sheng** [2]  **He Con-Gen** [2]  **Luo Cheng-Hong** [1]  **He Ji-Zhou** [1]

[1] Department of Physics, Nanchang University, Jiangxi, China, 330047

[2] Department of Mathematics, Nanchang University, Jiangxi, China, 330047

**Abstract**  We have investigated the Holmberg effect of Main galaxy pairs of the SDSS Data Release 4 (SDSS4). It is found that except i-z color  the color indices between the two components of Main galaxy pairs clearly have significantly larger correlation coefficients. Further analyses also show that the Holmberg Effect of galaxies not only depends on the color indices but also on the morphological type for two components of pairs.

**Keywords**   Galaxy: fundamental parameters-- galaxies: interactions

## 1. Introduction

Galaxy pairs have long been the important subjects of astronomical research and among the most fascinating astronomical objects in the universe. In the process of galaxy evolution, galaxy interactions frequently occur, and cause a variety of interesting astrophysical phenomena so that they play a crucial role in determining galaxy properties. In many studies, close galaxy pairs often be defined as interacting and merging galaxies, and are used for studing galaxy interactions.

For paired galaxies, some properties of both components may be correlated. More than 40 years ago, Holmberg(1958) , using a sample of 32 galaxy pairs, discovered that the color of paired galaxies are closely correlated, the correlation coefficient between the two components of pairs for $B-V$ color indices is R = $+0.80 \pm 0.06$. This result was interpreted as reflecting a tendency for galaxies forming and evolving together to have similar types. Later studies (Tomov 1978, 1979; Demin, Dibay, & Tomov 1981; Demin et al. 1984; Reshetnikov 1998) tended to confirm this result, finding that typical linear correlation coefficients between the two components of paired galaxies for the $B-V$ color indices are R $\approx 0.8$. However, above studies used very small samples of paired galaxies, their pair samples contained fewer than 100 galaxy pairs. This made these results insignificant.

Allam et al.(2004) calculated the Holmberg Effect for 1479 merging pairs in the Sloan Digital Sky Survey Early Data Release (SDSS EDR; Stoughton et al. 2002). The linear correlation coefficients for different color indices were following:

$R_{u-g} = 0.20 \pm 0.03$

$R_{g-r} = 0.38 \pm 0.03$

$R_{r-i} = 0.19 \pm 0.03$

$R_{i-z} = 0.07 \pm 0.04$

where the ($1\sigma$) errors were calculated via 1000 bootstrap resamplings of the original data.

Supported by the National Science Foundation of China (10465003)

They noted that the level and the significance of the correlation depends on the color indices. It is highest and has the greatest significance ($>12\sigma$) for g-r indices (which is the closest SDSS analog to B-V), while it is lowest and has the least significance ($<2\sigma$) for i-z indices. They further confirmed that these measured correlations are due to a physical effect rather than due to systematic errors of measurements.

Although the Holmberg effect has been investigated in many studies, it is undoubtedly of interest to confirm its existence using larger pair sample. The SDSS Data Release 4 (Adelman-McCarthy et al. 2005) provides a much deeper and wider galaxy sample with well-measured photometric and spectroscopic properties than almost any other wide-area catalog to date. From it, we can produce the largest galaxy pair sample and make ideal statistical analyses.

When identifying galaxy pairs, Many previous works were based on the two-dimensional projected sky separation and galaxy diameter (Karachentsev 1972; Lambas et al. 2003; Allam et al. 2004). The galaxy pair catalog identified by these selection criteria were actually two-dimensional sample. In order to produce a real three-dimensional pair sample, we use three-dimensional cluster analysis and extract close galaxy pairs from the Main galaxy sample (Strauss et al. 2002) of SDSS Data Release 4.

Our paper is organized as follows. In section 2, we describe the data to be used. The cluster analysis and selection criteria are discussed in section 3 and section 4 respectively. In section 5, we calculate the linear correlation coefficients between two components of Main galaxy pairs for different color indices. Our main results and conclusions are summarized in section 6.

## 2. Data

The Sloan Digital Sky Survey (SDSS ) is one of the largest astronomical surveys to date. The completed survey will cover approximately 10000 square degrees. York et al. (2000) provided the technical summary of the SDSS. The SDSS observes galaxies in five photometric bands (u, g, r, i, z) centered at (3540, 4770, 6230, 7630, 9130Å). The imaging camera was described by Gunn et al. (1998), while the photometric system and the photometric calibration of the SDSS imaging data were roughly described by Fukugita et al. (1996), Hogg et al. (2001) and Smith et al. (2002) respectively. Pier et al. (2003) described the methods and algorithms involved in the astrometric calibration of the survey, and present a detailed analysis of the accuracy achieved. Many of the survey properties were discussed in detail in the Early Data Release paper (Stoughton et al. 2002). Galaxy spectroscopic target selection can be implemented by two algorithms. The primary sample (Strauss et al. 2002), referred to here as the MAIN sample, targets galaxies brighter than r < 17.77(r-band apparent Petrosian magnitude). The surface density of such galaxies is about 90 per square degree. This sample has a median redshift of 0.10 and few galaxies beyond z=0.25. The Luminous Red Galaxy (LRG) algorithm (Eisenstein et al. 2001) selects $\approx 12$ additional galaxies per square degree, using color-magnitude cuts in g, r and i to select galaxies to r< 19.5 that are likely to be luminous early-types at redshifts up to $\approx 0.5$.

The SDSS has adopted a modified form of the Petrosian (1976) system for galaxy photometry, which is designed to measure a constant fraction of the total light independent of the surface-brightness limit. The Petrosian radius $r_p$ is defined to be the radius where the local surface-brightness averaged in an annulus equals 20 percent of the mean surface-brightness interior to this annulus, i.e.

$$\frac{\int_{0.8r_p}^{1.25r_p} dr 2\pi \, rI(r)/[\pi(1.25^2-0.8^2)r^2]}{\int_0^{r_p} dr 2\pi \, rI(r)/[\pi r^2]} = 0.2$$

where I(r) is the azimuthally averaged surface-brightness profile. The Petrosian flux $F_p$ is then defined as the total flux within a radius of $2r_p$, $F_p = \int_0^{2r_p} 2\pi \, rdrI(r)$. With this definition, the Petrosian flux (magnitude) is about 98 percent of the total flux for an exponential profile and about 80 percent for a de Vaucouleurs profile. The other two Petrosian radii listed in the Photo output, $R_{50}$ and $R_{90}$, are the radii enclosing 50 percent and 90 percent of the Petrosian flux, respectively.

The SDSS sky coverage can be separated into three regions. Two of them are located in the north of the Galactic plane, one region at the celestial equator and another at high declination. The third lies in the south of the Galactic plane, a set of three stripes near the equator. Each of these regions covers a wide range of survey longitude.

In our work, we consider the Main galaxy sample. The data is download from the Catalog Archive Server of SDSS Data Release 4 (Adelman-McCarthy et al. 2005) by the SDSS SQL Search (with SDSS flag: bestPrimtarget=64) with high-confidence redshifts (Zwarning $\neq$ 16 and Zstatus $\neq$ 0, 1 and redshift confidence level: zconf>0.95) (http://www.sdss.org/dr4/). From this sample, we select 260928 Main galaxies in redshift region: $0.02 \leq z \leq 0.2$.

In calculating the distance we use a cosmological model with a matter density $\Omega_0 = 0.3$, cosmological constant $\Omega_\Lambda = 0.7$, Hubble's constant $H_0 = 100h \text{km} \cdot \text{s}^{-1} \cdot \text{Mpc}^{-1}$ with h=0.7.

## 3. Cluster analysis

Cluster analysis (Einasto et al. 1984), as a general method, has been widely applied to study the geometry of point samples in many fields. The key of this method is to separate the sample into individual systems by an objective, automatic procedure. Let us draw a sphere of radius R around each sample point (in our case, galaxy). If within this sphere there are other galaxies they are considered belonging to the same system. Call these close galaxies "friends". Now draw spheres around new neighbours and continue the procedure using the rule "any friend of my friend is my friend". When no more new neighbours or "friends" can be added, then the procedure stops and a system is identified. As a result, each system consists of either a single, isolated galaxy or a number of galaxies which at least have one neighbour within a distance not exceeding R.

## 4. Selection criteria

At small neighbourhood radii only close double and multiple galaxies, cores of groups and conventional clusters of galaxies will form systems, the rest being isolated single galaxies. Apparently, close double systems identified at small neighbourhood radii are good candidates for galaxy pair sample. To define close pairs we have to choose the proper neighbourhood radius. Because we do not have a good priori criterion for this radius, we will consider a certain range of neighbourhood radii first and then analyse the properties of systems forming in the sample.

Through this procedure we find the proper neighbourhood radius and then identify close pairs at this radius.

From the original CfA2 redshift survey, Barton et al. (2000) extracted a complete sample of 786 galaxies in pairs and N-tuples which were selected to have projected separations $r_p < 50h^{-1}$ kpc and velocity separations $\Delta V \leq 1000$ km/s. Lambas et al. (2003) studied galaxy pairs in the field selected from the 100 K public release of the 2dFgalaxy redshift survey. Galaxy pairs were selected by radial velocity ($\Delta V \leq 350$ km/s) and projected separation ($r_p \leq 100 \text{kpc}$) criteria.

In Zitelli et al. (2004)'s volume-limited sample of 84 isolated pairs of galaxies, The projected separation between pair members is $r_p < 200 \, h^{-1}$ kpc. Because we need to select close galaxy pairs, the range of above projected separation criteria can be selected as the analysis range of our neighbourhood radii. We analyse the clustering properties of Main galaxy sample in the neighbourhood radius range: $R = 60\text{kpc} \rightarrow 200\text{kpc}$, in order to find the appropriate neighbourhood radius to identify galaxy pairs. At radius $R = 60\text{kpc}$, only close double and triplet galaxies form systems (the number of triplet systems is 9, the number of galaxies in close double systems is 1374), the rest being isolated galaxies. These close double systems can be considered good galaxy pair candidates. With the growth of neighbourhood radii the number of close double and multiple systems rapidly increases. When the radius reaches $100\text{kpc}$, the multiple systems include 184 galaxies, the close double systems contain 3342 galaxies. It should be noticed that some of close double systems forming at radius $R = 60\text{kpc}$ now are included into multiple systems. Fig.1 illustrates the change of the number of galaxies containing in multiple systems with the growth of neighbourhood radii.

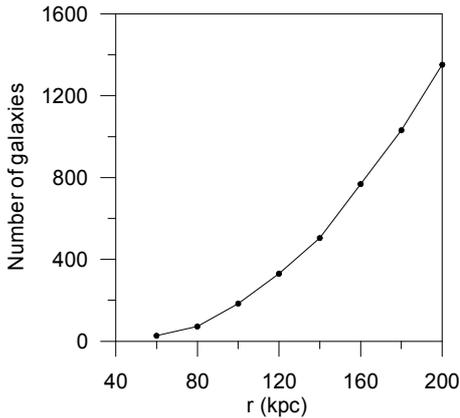 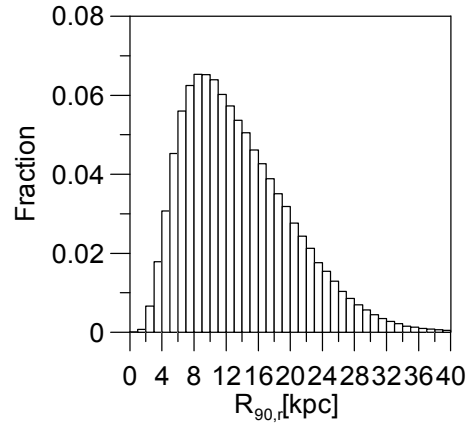

Fig.1 The change of the number of galaxies containing in multiple systems with the growth of neighbourhood radii.

Fig.2 The distribution of the r-band $R_{90}$ ($R_{90,r}$) of all galaxies for Main galaxy sample.

Fig.2 shows the distribution of the r-band $R_{90}$ ($R_{90,r}$) of all galaxies for Main galaxy sample. We notice that $R_{90,r}$ of few galaxies reaches to about 40 kpc. If we define 40kpc as the upper limit of galaxy radius size (if so, 80 kpc is minimum separation at which pairs can reliably be separated) and consider that at larger neighbourhood radius some of close double systems will merge into multiple systems, the neighbourhood radius $R \approx 100 \text{kpc}$ can be defined as appropriate neighbourhood radius to identify close galaxy pairs. Therefore, we select the close double systems identified at radii R=100 kpc as the galaxy pair sample, it contains 3342 galaxies (1.28 % of the total galaxy number).

SDSS observes spectra from targets, using a multi-object fiber spectrograph which can simultaneously observe 640 objects in a $3°$ diameter plate. Each plate can accommodate 640 fibers and fibers are assigned to targets, about 50 of these are reserved for calibration targets, leaving around 590 fibers for science targets. The fiber diameter is 0.2 mm ($3''$ on the sky). Because of the diameter of the cladding holding the optical fibers, adjacent fibers cannot be located more closely than $55''$ on the sky; it corresponds to a distance of about 120 kpc at z=0.10 ( the median redshift of Main galaxies). Roughly 8%-9% of selected objects are not observed for this reason, in regions covered by a single plate. Both members of a pair of objects closer than $55''$ can be observed spectroscopically only if they lie in the overlapping regions of adjacent plates, about 30% of the sky is covered by such overlaps (Stoughton et al. 2002). Apparently, this minimum fiber separation is a real hindrance for the study of the close pairs of SDSS. We hope that future survey can observe an even higher fraction of galaxies in close pairs by placing plate overlaps in regions where there is a high density of close pairs. If we study the large-scale distribution of pairs, above imcompleteness of the pair sample will be a large drawback. But when we study other properties (for example, the size, luminosity $M_r$ and concentration index ci) of pairs, the influence of this imcompleteness is not crucial.

To test our isolation criteria for our close pair sample, we have calculated the separation between each pair and its nearest neighboring galaxy. This separation is called "nearest-neighbor distance" of pairs and is defined as a parameter reflecting isolation property of pairs. We have analysed the distribution of "nearest-neighbor distance" of pairs for the pair sample. In our pair sample, the three–dimensional separation between both members of pairs is $R \leq 100 \text{kpc}$. We divide the range of "nearest-neighbor distance" of pairs into 5 subintervals:

$100 \text{kpc} < r_n \leq 200 \text{kpc}; 200 \text{kpc} < r_n \leq 300 \text{kpc}; 300 \text{kpc} < r_n \leq 400 \text{kpc}; 400 \text{kpc} < r_n \leq 500 \text{kpc}; r_n > 500 \text{kpc}$

and show the fraction of pairs in each subinterval as histograms. The whole argument range from 0 to 5. The first bin in the histogram is $100 \text{kpc} < r_n \leq 200 \text{kpc}$, the second $200 \text{kpc} < r_n \leq 300 \text{kpc}$, the third $300 \text{kpc} < r_n \leq 400 \text{kpc}$ and so on. Fig.3 shows the distribution of "nearest-neighbor distance" of pairs for the pair sample. The "nearest-neighbor distance" of 69.3% pairs is $r_n$>500kpc (>5 times separation between both members of pairs). Pairs of "nearest-neighbor distance" $100 \text{kpc} < r_n \leq 200 \text{kpc}$ (close to separation between both members of pairs) only are about 10.17%. Using the isolation criterion which requires pairs to have the "nearest-neighbor distance" $r_n>500$kpc. we construct a isolated galaxy pair sample. It contains 1158 pairs. Because redshift accuracy for the Main galaxy sample is 30km/sec rms, the two components of a pair have the same redshifts by our selection criteria.

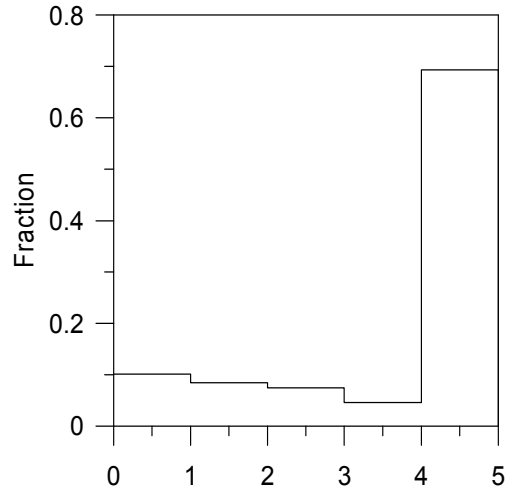

Fig.3　The distribution of "nearest-neighbor distance" of pairs for the pair sample.

## 5.　The Holmberg effect of Main galaxy pairs

Figure. 4 graphically illustrates the correlation between color indices of two components of 1158 pairs. The linear correlation coefficients for different color indices are following:

$$R_{u-g} = 0.514, R_{g-r} = 0.571, R_{r-i} = 0.417, R_{i-z} = 0.060.$$

The correlation coefficients apparently depends on the color indices. It is largest for g-r indices, while it is smallest for i-z indices. Allam et al.(2004) indicated that because most galaxies have very similar i-z colors (Fukugita, Shimasaku, & Ichikawa 1995), thus even small random photometric errors (< 0.05 mag) may tend to wash out any correlations of i-z color. We note that except i-z color the linear correlation coefficients for different color indices are apparently larger than Allam et al.(2004)'s results. Yee et al. (1995) defined three classes of pairs. Optical pairs were defined as galaxies that are close in angular projection. Physical pairs are objects which have similar redshifts, but their true separations may be larger than the projected distance limit. Of the physical pairs, there is a subset which consists of galaxies that have true separations smaller than the projected distance limit. These galaxies were referred to as "true" close pairs and formed the most likely merging population. Because Allam et al.(2004) identified pairs by the selection criteria of projected sky separation and galaxy diameter, their pair catalog may include some optical pairs or nonmerging associated pairs, while all of isolated pair sample are "true" close pairs and merging population. The lower significance of Allam et al.(2004)'s results are most likely due to contamination by paired galaxies that are either optical pairs or nonmerging associated pairs.

From the Main galaxy sample of SDSS Data Release 4, we randomly select 1158 pairs of galaxies and construc a random pair sample. Because of the flux-limited nature of the parent sample, our isolated pair sample only includes pairs with both components brighter than the flux limit , while those with the primary brighter than the limit but the secondary fainter than the limit are dropped. Missing secondaries are perhaps the most important source of incompleteness of a pair sample (Xu et al. 2004). This bias makes that most of pairs are composed of galaxies with magnitude close to the magnitude limit, these pairs are biased to have similar r magnitudes. In order to decrease the influence exerted by this bias, we have the random pair sample being affected by the bias which is exactly the same as that of the isolated pair sample. This requires that the random pairs are composed of galaxies with same redshifts, and the z distribution of the random pair sample being the same as that of isolated pair sample. We construct 10000 random pair samples and calculate the Holmberg Effect for random pair samples. The correlation coefficients for different color indices are following:

$$R_{u-g} = 0.177 \pm 0.038, R_{g-r} = 0.362 \pm 0.048, R_{r-i} = 0.265 \pm 0.070, R_{i-z} = 0.105 \pm 0.040.$$

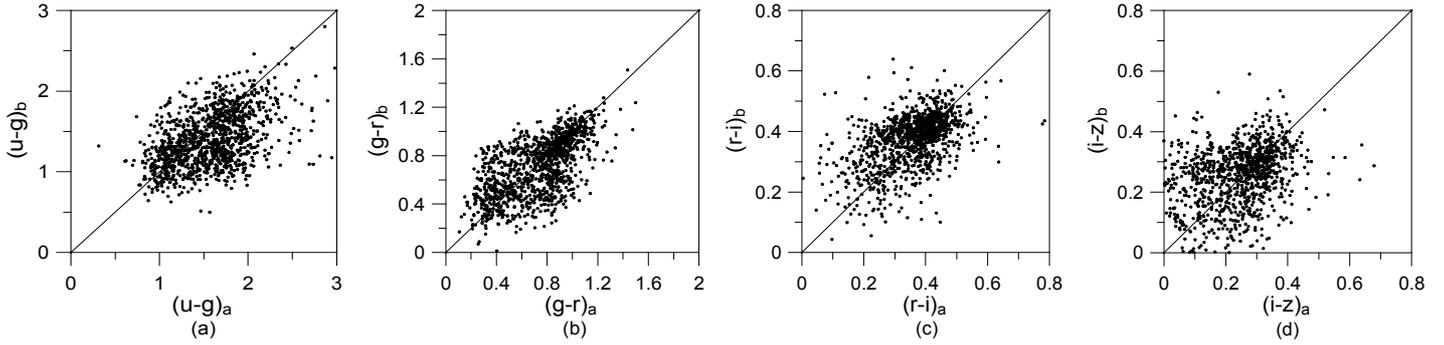

Fig.4 Correlation between color indices of two components of 1158 pairs. The subscript a refers to the fainter component and b to the brighter component. A line of 45° slope is also plotted. (a) for the u-g color, (b) for the g-r color, (c) for the r-i color, (d) for the i-z color.

Comparing the correlation coefficients of isolated pair sample with that of random pair samples, we find that except i-z color the color indices between the two components of pairs for isolated galaxy pair sample clearly have significantly larger correlation coefficients. This further confirms the Holmberg Effect. This correlation (color indices) suggests that both components of pairs have the same star composition.

Although apparently statistically significant, we can not make sure that this correlation is due to a physical effect. After all, it is quite possible that the basic properties of both galaxies in a close pair appear similar because the photometry of galaxy "a" is contaminated by the photometry of galaxy "b". Deng et al.(2005) performed the comparative studies of luminosity and size between both components of pairs for isolated Main galaxy pair sample from the SDSS Data Release 4, and found that there is no a tendency for paired galaxies to have similar luminosity and size. This demonstrates that such cross-contamination in the galaxy photometry may be insignificant, the color correlation between the two components of pairs is a physical effect.

There are attempts to classify SDSS galaxies into morphological classes through direct inspection of the galaxy images (Shimasaku et al. 2001; Nakamura et al. 2003). While such eye-ball classification should match the original Hubble morphological sequence, it is quite tedious and has so far been carried out only for about 1500 big galaxies in the SDSS. However, it has been suggested that some photometric and spectroscopic properties may be closely correlated with the morphological type, and so can be used as morphology indicators. For example, Shimasaku et al.(2001) showed that the concentration index $c_i = R_{90} / R_{50}$ can be used to separate early-type (E/S0) galaxies from late-type (Sa/b/c, Irr) galaxies. Using about 1500 galaxies with eye-ball classification, Nakamura et al. (2003) confirmed that $c_i = 2.86$ separates galaxies at S0/a with a completeness of about 0.82 for both late and early types. According to types of both components in each pair, we separate the isolated pair sample into three subsamples, and respectively refer to early+early subsample, late+late subsample and early+late subsample. Early+early subsample includes 108 pairs in which both components are early-type galaxies ($c_i \geq 2.86$), late+late subsample contains 693 pairs in which both components are late -type galaxies ($c_i < 2.86$), and early+late subsample contains 357 pairs in which both components are different types. For different pair subsamples, we also construct 10000 corresponding random pair samples being affected by the bias which is exactly the same as that of different pair subsamples. Both components of random pairs have types being the same as that of different pair subsamples (for example, random pairs of early+early subsample are early+early galaxies). Fig.5, Fig.6 and Fig.7 illustrate correlation between color indices of two components for 108 early+early pairs, 693 late+late pairs and 357 early+late pairs respectively. We further calculate the linear correlation coefficients of different color indices for different pair subsamples and corresponding random pair samples. The results are listed in Table 1. We find that early+early pairs do not exhibit statistically significant Holmberg Effect, while late+late pairs and early+late pairs clearly have statistical correlation coefficients between the two components of pairs except i-z color. This indicate that the Holmberg Effect of galaxies not only depends on the color indices but also on the morphological type for two components of pairs.

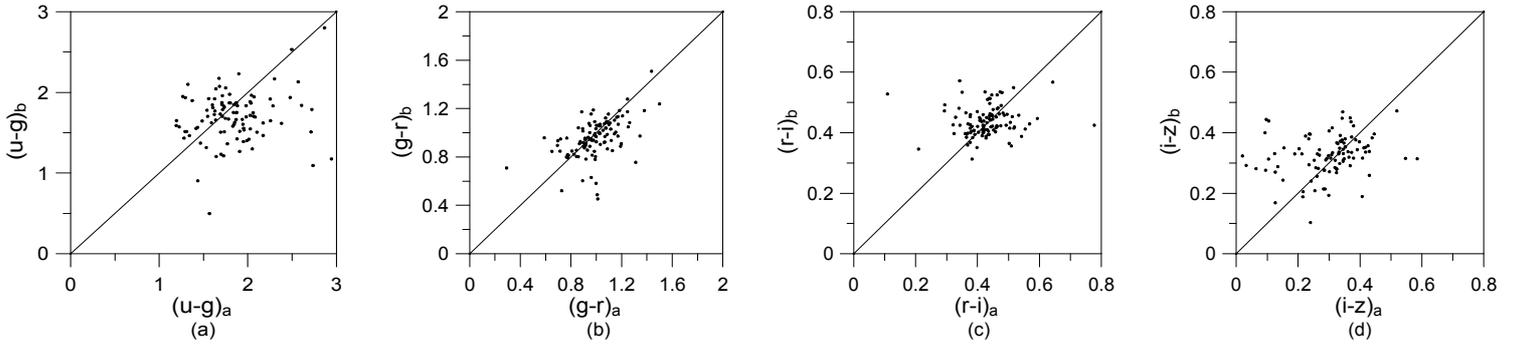

Fig.5  Correlation between color indices of two components of 108 early+early pairs. The subscript a refers to the fainter component and b to the brighter component. A line of 45° slope is also plotted. (a) for the u-g color, (b) for the g-r color, (c) for the r-i color, (d) for the i-z color.

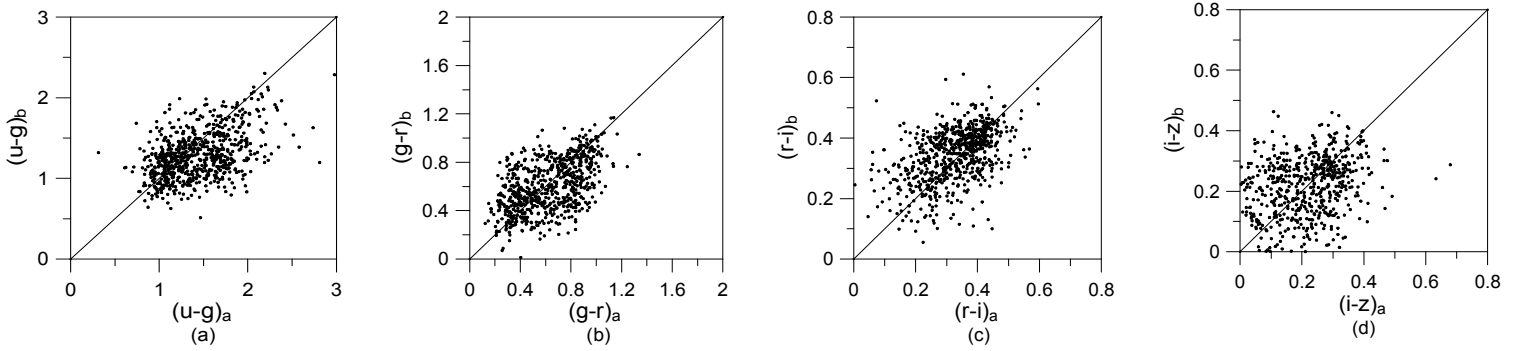

Fig.6　Correlation between color indices of two components of 693 late+late pairs. The subscript a refers to the fainter component and b to the brighter component. A line of 45° slope is also plotted. (a) for the u-g color, (b) for the g-r color, (c) for the r-i color, (d) for the i-z color.

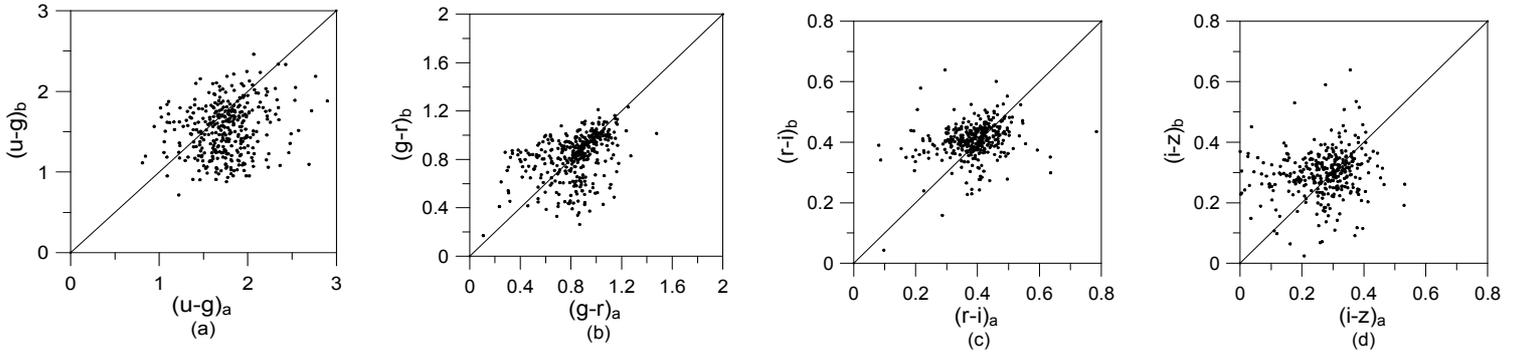

Fig.7　Correlation between color indices of two components of 357 early+late pairs. The subscript a refers to the fainter component and b to the brighter component. A line of 45° slope is also plotted. (a) for the u-g color, (b) for the g-r color, (c) for the r-i color, (d) for the i-z color.

## 6. Summary

Using the MAIN galaxy data from the SDSS Data Release 4 (SDSS4), we have identified close galaxy pairs by three-dimensional cluster analysis, and investigate the Holmberg effect of Main galaxy pairs. The Main galaxy sample is limited in the redshift region: $0.02 \leq z \leq 0.2$ and contains 260928 Main galaxies. Galaxy pairs are identified at neighbourhood radius $R = 100\text{kpc}$.

Using the isolation criterion which requires pairs to have the "nearest-neighbor distance" $r_n$>500kpc. we finally construct a isolated galaxy pair sample. It contains 1158 pairs. Comparing the correlation coefficients of isolated pair sample with that of random pair samples, we find that except i-z color the color indices between the two components of pairs for isolated galaxy pair sample clearly have significantly larger correlation coefficients. Further analyses also show that the Holmberg Effect of galaxies not only depends on the color indices but also on the morphological type for two components of pairs.

Table1: The correlation coefficients of the color indices between the two components of pairs for different pair subsamples and corresponding random pair samples

| The pair subsamples | The correlation coefficients of different pair subsamples | The correlation coefficients for corresponding random pair samples |
|---|---|---|
| Early+early pairs | $R_{u-g} = 0.240$<br>$R_{g-r} = 0.507$<br>$R_{r-i} = 0.369$<br>$R_{i-z} = 0.081$ | $R_{u-g} = 0.145 \pm 0.111$<br>$R_{g-r} = 0.562 \pm 0.086$<br>$R_{r-i} = 0.212 \pm 0.110$<br>$R_{i-z} = 0.084 \pm 0.107$ |
| Late+late pairs | $R_{u-g} = 0.474$<br>$R_{g-r} = 0.458$<br>$R_{r-i} = 0.365$<br>$R_{i-z} = 0.027$ | $R_{u-g} = 0.145 \pm 0.049$<br>$R_{g-r} = 0.270 \pm 0.038$<br>$R_{r-i} = 0.249 \pm 0.063$<br>$R_{i-z} = 0.076 \pm 0.051$ |
| Early+late pairs | $R_{u-g} = 0.648$<br>$R_{g-r} = 0.380$<br>$R_{r-i} = 0.235$<br>$R_{i-z} = 0.131$ | $R_{u-g} = 0.060 \pm 0.062$<br>$R_{g-r} = 0.128 \pm 0.046$<br>$R_{r-i} = 0.140 \pm 0.062$<br>$R_{i-z} = 0.025 \pm 0.054$ |

Funding for the creation and distribution of the SDSS Archive has been provided by the Alfred P. Sloan Foundation, the Participating Institutions, the National Aeronautics and Space Administration, the National Science Foundation, the U.S. Department of Energy, the Japanese Monbukagakusho, and the Max Planck Society. The SDSS Web site is http://www.sdss.org/.

The SDSS is managed by the Astrophysical Research Consortium (ARC) for the Participating Institutions. The Participating Institutions are The University of Chicago, Fermilab, the Institute for Advanced Study, the Japan Participation Group, The Johns Hopkins University, Los Alamos National Laboratory, the Max-Planck-Institute for Astronomy (MPIA), the Max-Planck-Institute for Astrophysics (MPA), New Mexico State University, University of Pittsburgh, Princeton University, the United States Naval Observatory, and the University of Washington.